\documentclass[12pt]{article}
\usepackage{fancybox}
\usepackage{amsmath,amssymb,amsfonts}
\usepackage{bm}
\usepackage[dvipdfmx]{graphicx}
\allowdisplaybreaks
\title{\bf Singular Gauge Transformation \\and the Erler-Maccaferri Solution \\
in Bosonic Open String Field Theory}
\author{Akitsugu {\sc Miwa}\thanks{\tt miwa.akitsugu@nihon-u.ac.jp}
 and Kazuhiro {\sc Sugita}\thanks{\tt sugita@phys.cst.nihon-u.ac.jp}\\[4mm]
{\it Department of Physics, College of Science and Technology,}\\
{\it Nihon University, 1-8-14, Kanda-Surugadai,}\\
{\it Chiyoda-ku, Tokyo, 101-8308, Japan}\\ 
}
\date{}
\makeatletter
  
  \@addtoreset{equation}{section}
\makeatother
\begin{document}
\maketitle
\vspace{2cm}
\begin{abstract}
We study candidates of the multiple-brane solutions of bosonic open string field theory.
They are constructed by performing a singular gauge transformation $n$ times for the Erler-Maccaferri solution.
We check the EOM in the strong sense,
and find that it is satisfied only when we perform the gauge transformation once.
We calculate the energy for that case and obtain a support that the solution is a multiple-brane solution.
We also check the tachyon profile for a specific solution
which we interpret as describing a D24-brane placed on a D25-brane.
\\
\end{abstract}

\vspace{8cm}

\thispagestyle{empty}
\setcounter{page}{0}
\newpage

\tableofcontents
\section{Introduction}

Since Schnabl constructed his analytic solution \cite{Schnabl:2005gv}, there has been remarkable progress 
in our understanding of analytic solutions in bosonic open string field theory \cite{Witten:1985cc}.
This includes several descriptions of backgrounds with multiple D-branes, i.e. multiple-brane solutions.

The first description uses singular gauge transformations \cite{Okawa:2006vm,Ellwood:2009zf,Murata:2011ex,Erler:2012qn,Erler:2012qr},
and building blocks in this approach are only string fields $K$, $B$ and $c$ \cite{Okawa:2006vm}.
The singular gauge transformation used in the paper \cite{Murata:2011ex} is the inverse of the one which is used in the construction of 
the ``simple'' tachyon vacuum solution \cite{Erler:2009uj} in the pure-gauge form \cite{Okawa:2006vm}.
The energy of the solution depends on how many times of the gauge transformation are
performed to construct the solution 
\cite{Murata:2011ex}.
However in \cite{Hata:2011ke}, it was shown that this type of multiple-brane solutions 
does not satisfy the equation of motion (EOM) in the strong sense except for the double-branes, the single-brane and the no brane cases.\footnote{
See also the recent paper \cite{Hata:2015aoh} for the study of the tachyon fluctuations around the solutions.
}
In that paper, such a result is derived by introducing a regularization \cite{Murata:2011ex,Hata:2011ke,Erler:2011tc}
for the singular string field $1/K$.\footnote{In this paper we only consider the singularity at $K=0$.}
The second type of solution is constructed by Erler and Maccaferri \cite{Erler:2014eqa}, i.e. the Erler-Maccaferri solution (the EM solution).
It uses the string fields originating from the insertion of the boundary condition changing operators (BCCOs) \cite{Kiermaier:2010cf}
in addition to $K$, $B$ and $c$.
Since the different boundary conformal field theories (BCFTs) correspond to different string backgrounds,
this type of solution is capable of describing backgrounds other than the perturbative vacuum. 

In this paper, by combining the above two descriptions together,
we give another type of candidates of multiple-brane solutions.
To be more precise, we perform the singular gauge transformation $n$ times for the EM solution.
Since the resulting string fields, just as the double-brane solution, include the factor $1/K$, 
we use the regularization of \cite{Murata:2011ex,Hata:2011ke,Erler:2011tc}.
By checking the EOM in the strong sense carefully,
we find that the calculations essentially reduce to the ones 
in the case without BCCOs \cite{Hata:2012cy}, 
and that the EOM in the strong sense is not satisfied
if we perform the singular gauge transformation more than once for the EM solution.
Recalling that, in the case without BCCOs, the effect of the singular gauge transformation
is essentially to increase the energy by one unit of the D25-brane's energy, 
one may expect that our solution includes a D25-brane coming from the gauge transformation 
in addition to the original D-brane existing in the EM solution.
Indeed, as we will see, the energy of the solution is given by the summation of the energy of these branes.
We also study the profile of the tachyon field in the case when the BCCOs are 
taken to be the ones which change the Neumann boundary condition to the Dirichlet boundary condition
and obtain further support for the above expectation.

This paper is organized as follows.
In section 2, we review the two types of multiple-brane solutions and we present our candidates for the solutions.
In section 3, we check the EOM in the strong sense and also calculate the energy.
In section 4, as a concrete example of our solution, we choose the EM solution describing a D24-brane 
and calculate the tachyon profile.
Section 5 is devoted to conclusions. 

\section{Multiple-Brane Solutions}
Our candidates are constructed by performing the singular gauge transformation for the EM solution.
Here, the gauge transformation is the one which transforms the perturbative vacuum 
to the double-brane solution in the pure-gauge form.
In this section, we first review the pure-gauge-form solutions 
and next the EM solution,  and then we give our candidates.

\subsection{Pure-Gauge-Form Solutions}
\label{Pure-Gauge}
The pure-gauge-form solutions are written by the string fields $K$, $B$ and $c$ \cite{Okawa:2006vm},\footnote{
We use the same convention as \cite{Okawa:2006vm}, which is called the right handed convention.
} which are defined in the sliver frame \cite{Rastelli:2000iu}. 
These string fields satisfy the following $KBc$ algebra,
\begin{equation}
\begin{array}{ccc}
[K,B]=0, & \quad B^2=c^2=0, \quad & \{B,c\}=1,\\
QK=0, & QB=K, & Qc=c\partial c.
\end{array}
\end{equation}
Here, $Q$ is BRST operator and we define $\partial c \equiv [K,c]$.
In the bosonic open string field theory, the gauge transformation is given by
\begin{align}
\varphi \to V^{-1}(Q+\varphi)V.
\end{align}
When we restrict the gauge parameter $V$ to the form
\begin{equation}
V=Bc+cBg(K),
\end{equation}
where $g(K)$ is a function of the string field $K$.
The string field $\Psi$ in the pure-gauge-form is given by
\begin{equation}
\Psi=V^{-1}QV=cBKg^{-1}c(1-g).
\end{equation}
Here, the inverse of $V$ is $V^{-1} = Bc+cBg(K)^{-1}$.
Since it is pure gauge, it formally satisfies the following EOM algebraically
\begin{equation}
Q\Psi+\Psi^2=0, \label{QP+PP=0}
\end{equation}
but for the same reason, it is gauge equivalent to the perturbative vacuum, $\Psi_1\equiv0$, 
unless the gauge transformation is singular in some sense.
Then if one wants a solution other than the perturbative vacuum, a singular gauge transformation is required.

The ``simple'' tachyon vacuum solution \cite{Erler:2009uj} is given by
\begin{equation}
U^{-1}QU=-cB(1-K)c\frac{1}{1-K} \equiv \Psi_0,
\end{equation}
where 
\begin{equation}
U=Bc+cBG(K),\quad G(K)\equiv\frac{-K}{1-K}.\label{UI}
\end{equation}
The energy of this solution is lower than the perturbative vacuum by that of a D25-brane.
Since this solution is not the perturbative vacuum, the gauge transformation by the gauge parameter \eqref{UI} is singular,
while the solution $\Psi_0$ itself is regular.

The inverse gauge transformation gives
\begin{equation}
UQU^{-1}
=
cBKGc(1-G^{-1})
=
cB\frac{K^2}{1-K}c\frac{1}{-K}\equiv \Psi_2.
\label{psi2}
\end{equation}
This is a double-brane solution reproducing twice the energy of the D25-brane\footnote{
Here and in the following we express the energy in the form ``$E(\Psi_0) + \cdots$''
in order to clarify the excitations above the tachyon vacuume $\Psi_0$.
}
\begin{equation}
E(\Psi_2)
=
{\rm Tr}
\Big{[}
	\frac{1}{2}\Psi_2 Q\Psi_2+\frac{1}{3}{\Psi_2}^3
\Big{]}
=
\frac{1}{2\pi^2}
=
E(\Psi_{0})+2\times\frac{1}{2\pi^2}.
\label{E2}
\end{equation}
It satisfies the EOM in the strong sense which is defined as %i.e. %the inner product between the solution itself and the EOM vanishes:
\begin{equation}
{\rm EOMS}(\Psi_2)
\equiv
{\rm Tr}\Big{[} \Psi_2(Q\Psi_2+{\Psi_2}^2)\Big{]}
=
0.
\label{EOMS2}
\end{equation}
There could be also other solutions which are constructed by using $U^{-1}$ more than once, i.e.
\begin{equation}
\Psi_n
=
U^{n-1}QU^{-(n-1)}
=
cBK{G}^{n-1}c(1-{G}^{-(n-1)}),
\quad n\ge 3.
\end{equation}
However, it is shown that they
do not satisfy the EOM in the strong sense \cite{Hata:2011ke,Hata:2012cy}.

Let us explain the calculation of \eqref{E2} and \eqref{EOMS2} more precisely.
As mentioned above, 
the evaluation of \eqref{E2} and \eqref{EOMS2} 
could suffer from the singularities.
A possible origin of the singularity is the factor $1/K$ 
in \eqref{psi2} which is singular at $K=0$
and it needs some regularization.
A standard regularization scheme in literature is the so called ``$K_\epsilon$ regularization",\footnote{
See, \cite{Masuda:2012cj} for another regularization scheme.
}
in which we replace each $K$ in the solution of \eqref{QP+PP=0} with $K_\epsilon$ defined by
\begin{equation}
K_\epsilon\equiv K-\epsilon, \quad \epsilon>0.
\end{equation}
After this regularization, the algebra among $K_\epsilon$, $B$ and $c$ is given by
\begin{equation}
\begin{array}{ccc}
[K_\epsilon,B]=0, & \quad 
B^2=c^2=0, \quad & 
\{B,c\}=1,\\
QK_\epsilon=0, &
QB=K_\epsilon+\epsilon, &
Qc=c\partial c,
\end{array}
\end{equation}
where $\partial c = [K, c] = [ K_\epsilon, c ]$.

Next one may rewrite $1/(1-K_\epsilon)$ and the factors $K_\epsilon$ both in the numerator and in the denominator
in terms of the wedge state% by introducing the Schwinger parameters
\begin{align}
\frac{1}{1-K_\epsilon } 
& =
\int_0^\infty dx\,
e^{-(1+\epsilon)x}\Omega^{x},\label{x}\\[0.5mm]
K_\epsilon 
& =
\frac{\partial}{\partial y} e^{-\epsilon y }\Omega^{y}|_{y=0}, \\[0.5mm]
{1 \over K_\epsilon} 
& = - \int_0^\infty dz\,
e^{-\epsilon z} \Omega^{z}.
\label{z}
\end{align}
Here, $\Omega$ is defined as $\Omega\equiv e^{K}$ and $x$, $y$ and $z$ are the widths in the sliver frame.
Then \eqref{E2} and \eqref{EOMS2}
can be evaluated 
by 
%interpreting these Schwinger parameters as the widths on the world sheet and 
using CFT correlators.

In the following sections we use the notation 
$[\cdot]_\epsilon$ for the $K_\epsilon$ regularization in which $K$ inside the square bracket with subscript $\epsilon$ 
is replaced with $K_\epsilon$
\begin{equation}
[f(K,B,c)]_\epsilon
=
f(K_\epsilon,B,c) .
\end{equation}
In this notation, precise meanings of \eqref{E2} and \eqref{EOMS2} are
\begin{equation}
\lim_{\epsilon \to 0} E\big( [ \Psi_2 ]_\epsilon \big)
=
{1\over 2 \pi^2} , \quad 
\lim_{\epsilon \to 0} {\rm EOMS}\big( [\Psi_2 ]_\epsilon \big)
=
0 .
\end{equation}

\subsection{Erler-Maccaferri Solution}
Next, we review the EM solution \cite{Erler:2014eqa}.
The form of the solution is given as follows:
\begin{align}
\Psi_{{}^{\rm EM}}^{\rm a} 
& \equiv
\Psi_0+\Sigma_{\rm L}^{\rm a}(-\Psi_0)\Sigma_{\rm R}^{\rm a}\notag\\
&=
-cB(1-K)c\frac{1}{1-K}+cB(1-K)\sigma_{\rm L}^{\rm a}\frac{1}{1-K}\sigma_{\rm R}^{\rm a}(1-K)c\frac{1}{1-K},
\label{EM}
\end{align}
where $\Sigma^{\rm a}_{\rm L}$ and $\Sigma^{\rm a}_{\rm R}$ are defined by\footnote{$\Sigma^{\rm a}_{\rm L,R}$, in this paper, 
are in the non-real forms, while the original solution uses the real forms \cite{Erler:2014eqa}.}
\begin{equation}
\Sigma_{\rm L}^{\rm a}
\equiv
Q_0(\sigma_{\rm L}^{\rm a}A_0), 
\quad \Sigma_{\rm R}^{\rm a}
\equiv 
Q_0(\sigma_{\rm R}^{\rm a}A_0).
\label{Sigma-Sigma}
\end{equation}
The string fields $\sigma_{\rm L}^{\rm a}$ 
and $\sigma_{\rm R}^{\rm a}$ are made by inserting the BCCOs on the string world sheet boundary, 
which change the boundary condition for BCFT$_0$\footnote
{
Here, the boundary condition of BCFT$_{0}$ is the one corresponding to the perturbative vacuum, i.e. D25-brane.} 
into the one for some other BCFT$_{\rm a}$ and vice versa.
$Q_0$ is the shifted kinetic operator around the tachyon vacuum, $Q_0\varphi \equiv Q\varphi+[\Psi_0,\varphi]$, 
and $A_0 \equiv BG/K$ is its homotopy operator.
Note that if we replace each $\sigma^{\rm a}_{\rm L}$
and $\sigma^{\rm a}_{\rm R}$ 
in \eqref{EM} and 
\eqref{Sigma-Sigma} with an identity string field $1$,
%i.e. if there is no change in the boundary conditions,
then both $\Sigma_{\rm L}^{\rm a}$ and 
$\Sigma_{\rm R}^{\rm a}$ are equal to $1$, 
and the solution $\Psi^{\rm a}_{{}^{\rm EM}}$
becomes perturbative vacuum:
\begin{equation}
\Psi^{\rm a}_{{}^{\rm EM}}\big|_{\sigma_{\rm L, R}^{\rm a}=1}
=
\Psi_1
=
0.
\label{Psi=0}
\end{equation}

The $KBc$ algebra is now generalized to include the 
algebraic relations among $K$, $B$, $c$ and $\sigma_{\rm L,R}^{\rm a}$.
The additional relations are 
\begin{align}
[B,\sigma_{\rm L,R}^{\rm a}]
=
[c,\sigma_{\rm L,R}^{\rm a}]
=
[B,\partial \sigma_{\rm L,R}^{\rm a}]
=
[c,\partial \sigma_{\rm L,R}^{\rm a}]
=
0,~
\sigma_{\rm R}^{\rm a}\sigma_{\rm L}^{\rm a}
=
1,~
\sigma_{\rm L}^{\rm a}\sigma_{\rm R}^{\rm a}
=
g_{\rm a},
\label{algghBCCO}
\end{align}
where $\partial \sigma^{\rm a}_{\rm L, R} = [ K, \sigma^{\rm a}_{\rm L, R}]$ and 
$g_{\rm a}$ is the disk partition function of BCFT$_{\rm a}$.
The energy of this solution is given by
\begin{equation}
E(\Psi^{\rm a}_{{}^{\rm EM}})
=
\frac{1}{2\pi^2}(-1+g_{\rm a})
=
E(\Psi_0) + {g_{\rm a} \over 2 \pi^2}.
\end{equation}
Here, since the solution does not include the factor $1/K$, no regularization are needed.

The EM solution describing the background with multiple D-branes can be constructed 
by using the orthogonal BCCOs satisfying the relation
\begin{equation}
\sigma_{\rm R}^{\rm a}\sigma_{\rm L}^{\rm b}
=
0, \quad
\sigma_{\rm L}^{\rm a}\sigma_{\rm R}^{\rm b}
=
0,
\quad {\rm a} \neq {\rm b}.
\label{sigmaab}
\end{equation}
In the case with two D-branes, the solution can be written down as follows:
\begin{align}
\Psi^{\rm a+b}_{{}^{\rm EM}}
=
\Psi_{0}
+
\Sigma_{\rm L}^{\rm a}(-\Psi_0)\Sigma_{\rm R}^{\rm a}
+
\Sigma_{\rm L}^{\rm b}(-\Psi_0)\Sigma_{\rm R}^{\rm b}.
\label{Psi-EM-a+b}
\end{align}
The energy of this solution is given by
\begin{align}
E(\Psi^{\rm a+b}_{{}^{\rm EM}})
=
\frac{1}{2\pi^2}(-1+g_{\rm a}+g_{\rm b})
=
E(\Psi_0) + { g_{\rm a} \over 2\pi^2} + { g_{\rm b} \over 2 \pi^2}
.
\end{align}
The generalization to the case with more D-branes should be straightforward.
The matrix structure for the fluctuations around the background 
is investigated in \cite{Kishimoto:2014yea}.

\subsection{Our Solutions}
In this subsection, we discuss the string fields constructed by 
performing the gauge transformation $n$ times
for the EM solution $\Psi^{\rm a}_{{}^{\rm EM}}$.
By performing the gauge transformation with the parameter $U^{-1}=Bc+cB(-K/(1-K))^{-1}$ once,
we obtain  
\begin{align}
\Psi_{{}^{\rm EM}{}^{+1}}^{\rm a}& \equiv U(Q+\Psi_{{}^{\rm EM}}^{\rm a})U^{-1} \notag \\
&=
U\Sigma_{\rm L}^{\rm a}(-\Psi_0) \Sigma_{\rm R}^{\rm a} U^{-1}\notag \\
&=
cBK\sigma_{\rm L}^{\rm a}\frac{1}{1-K}\sigma_{\rm R}^{\rm a}Kc\frac{1}{-K}, 
\label{a+1}
\end{align}
while performing it $n$ times gives 
\begin{align}
\Psi_{{}^{\rm EM}{}^{+n}}^{\rm a}
&=
U^{n}(Q+\Psi_{{}^{\rm EM}}^{\rm a})U^{-n}\notag \\
&=
\Psi_{n}
+cBKG^{n-1}\sigma_{\rm L}^{\rm a}\frac{1}{1-K}\sigma_{\rm R}^{\rm a}
(
	-c+\partial c \frac{1}{-K}G^{-(n-1)}
)\notag \\
&\equiv
\Psi_{n}+\Phi_n^{\rm a}.
\label{a+n}
\end{align}
Although these string fields formally satisfy the EOM \eqref{QP+PP=0},
it dose not mean that they satisfy the EOM in the strong sense.
Then we may call them as candidates for the solutions.
We will check the EOM in the strong sense and 
compute the energy of the solutions in the next section
by using $K_{\epsilon}$ regularization.
We note that if we replace $\sigma^{\rm a}_{\rm L}$ and
$\sigma^{\rm a}_{\rm R}$  in \eqref{a+1} 
with the identity string field 1, 
it becomes the double-brane solution \eqref{psi2}, i.e. 
$\Psi_{{}^{\rm EM}{}^{+1}}^{\rm a}|_{\sigma^{\rm a}_{\rm L,R} = 1}=\Psi_2$.
This can be seen also from the first line of \eqref{a+1},
since this replacement makes 
$\Psi^{\rm a}_{{}^{\rm EM}}|_{\sigma^{\rm a}_{\rm L,R} = 1} = 0$ as \eqref{Psi=0}. 
Then from the first line of \eqref{a+n}, it is 
also easy to see the general relation
\begin{equation}
\Psi_{{}^{\rm EM}{}^{+n}}^{\rm a}|_{\sigma^{\rm a}_{\rm L,R} = 1}
=
U^n Q U^{-n}
=
\Psi_{n+1}.
\label{Psi=Psi}
\end{equation}

As discussed in section\,\ref{Pure-Gauge}, 
the gauge transformation of $\Psi_1$ with the parameter $U^{-1}$ increases the number of the D25-brane by 1 and gives the double-brane solution $\Psi_2$.
Then one might expect that \eqref{a+1} describes the background with two D-branes, i.e. 
the D-brane originally exists in the EM solution and the additional D25-brane corresponding to the gauge transformation $U^{-1}$. 
We will provide a support for this observation by checking the EOM in the strong sense and calculating the energy of the solution in the following sections.
As for $\Psi_{{}^{\rm EM}{}^{+n}}^{\rm a}$ with $n\geq 2$, we show that the EOM in the strong sense 
is not generally satisfied.

\section{EOM in the Strong Sense and Energy of the Solution}
%In this section we study properties of our candidates and
%provide some supports for the expectation for our solution being the multiple-brane solution.
\subsection{EOM in the Strong Sense}
We start by checking the EOM in the strong sense
\begin{equation}
{\rm EOMS}(\Psi) \equiv {\rm Tr}[\Psi(Q\Psi+\Psi^2)] = 0
\end{equation}
for $\Psi = \Psi_{{}^{\rm EM}{}^{+n}}^{\rm a} $ which is given in \eqref{a+n}.
%Here, we have introduced the notation ${\rm EOMS}(\Psi)$ for the sake of later convenience.
Since these string fields include $1/K$ factor, we use the $K_\epsilon$ regularization.

Let us assume $\Psi$ to be any formal solution of the 
algebraic equation $Q \Psi + \Psi^2 = 0$ which is constructed from
the building blocks $K$, $B$, $c$ and $\sigma_{\rm L,R}^{\rm a}$.
Then, the following equation holds:
\begin{equation}
[Q\Psi  ]_\epsilon +  [\Psi ]_\epsilon^2=0,
\end{equation}
since the algebraic relations among the building blocks are 
not changed by the regularization, i.e. by the replacement $K \to K_\epsilon$.
We also have the following equations:
\begin{equation}
Q[K]_\epsilon = [QK]_\epsilon , \quad 
Q[B]_\epsilon = [QB]_\epsilon + \epsilon, \quad 
Q[c] _\epsilon = [Qc]_\epsilon , \quad 
Q[\sigma_{\rm L,R}^{\rm a} ]_\epsilon = [ Q \sigma_{\rm L,R}^{\rm a} ]_\epsilon .
\end{equation} 
By using these equations, we have
\begin{equation}
Q [ \Psi ]_\epsilon + [ \Psi ]_\epsilon{}^2  
= 
Q [ \Psi ]_\epsilon
-
[Q \Psi ]_\epsilon
=
\epsilon {\partial \over \partial B} [\Psi ]_\epsilon.
\end{equation}
Then taking the inner product with $[\Psi]_\epsilon$, we obtain 
\begin{equation}
 {\rm EOMS}( [\Psi]_\epsilon ) = {\rm Tr} \Big[ [\Psi]_\epsilon  \epsilon {\partial \over \partial B} [ \Psi ]_\epsilon \Big].
\label{eomsn}
\end{equation}
We apply this formula to $\Psi_{{}^{\rm EM}{}^{+n}}^{\rm a} = \Psi_n + \Phi_n^{\rm a}$ and $\Psi_n$,
and consider their difference
\begin{align}
& {\rm EOMS}\big{(}
	[\Psi_{{}^{\rm EM}{}^{+n}}^{\rm a}]_\epsilon
\big{)}
-{\rm EOMS}\big{(}[\Psi_n]_\epsilon\big{)} \notag \\
& \qquad =  
{\rm Tr}\Big{[} 
	( 
		[\Psi_n]_\epsilon + [\Phi_n^{\rm a}]_\epsilon 
	)  
	\epsilon {\partial \over \partial B}
	(
	 [\Psi_n]_\epsilon +  [\Phi_{n}^{\rm a}]_\epsilon
	)
\Big{]}
-
{\rm Tr} \Big[ 
	[\Psi_n]_\epsilon \epsilon {\partial \over \partial B} [\Psi_n]_\epsilon 
\Big].
\label{eoms}
\end{align}
Since each $\Phi_n^{\rm a}$ contains two BCCOs, the right hand side (RHS) of \eqref{eoms} seems to be 
composed of terms with two BCCOs and also four BCCOs. 
The explicit form of the term with four BCCOs is as follows:
\begin{align}
&{\rm Tr}\Big{[}[\Phi^{\rm a}_n]_\epsilon
(
	\epsilon {\partial \over \partial B}[\Phi_{n}^{\rm a}]_\epsilon
)\Big{]}\notag\\
&=
-\epsilon{\rm Tr}\Big{[}
	cBK_\epsilon G_\epsilon^{n-1} \sigma_{\rm L}^{\rm a}\frac{1}{1-K_\epsilon}\sigma_{\rm R}^{\rm a} 
	\partial c \frac{1}{-K_\epsilon} G_\epsilon^{-(n-1)} c K_\epsilon G_\epsilon^{n-1}
	\sigma_{\rm L}^{\rm a} \frac{1}{1-K_\epsilon} \sigma_{\rm R}^{\rm a} 
	\partial c \frac{1}{-K_\epsilon}G_\epsilon^{-(n-1)}
\Big{]}.
\label{EOMiSSa+n}
\end{align}
Here, we abbreviate as $[ G ]_\epsilon \to G_\epsilon$ in order to avoid ugly expressions. 
Now let us show that this term 
can be rewritten into the summation of terms with two BCCOs.
We first rewrite \eqref{EOMiSSa+n} by using the relation 
$cK_\epsilon G_\epsilon^{n-1} = [c,K_\epsilon G_\epsilon^{n-1}] + K_\epsilon G_\epsilon^{n-1}c $ as
\begin{align}
\eqref{EOMiSSa+n}
=
-\epsilon &{\rm Tr}\Big{[} 
	[c,K_\epsilon G_\epsilon^{n-1}] B
	\sigma_{\rm L}^{\rm a}\frac{1}{1-K_\epsilon}\sigma_{\rm R}^{\rm a}
	\partial c \frac{1}{-K_\epsilon} G_\epsilon^{-(n-1)} \notag\\
&\qquad
	\times [c,K_\epsilon G_\epsilon^{n-1}]
	\sigma_{\rm L}^{\rm a}\frac{1}{1-K_\epsilon}\sigma_{\rm R}^{\rm a}
	\partial c \frac{1}{-K_\epsilon} G_\epsilon^{-(n-1)}
 \Big{]}\notag\\
+
\epsilon &{\rm Tr}\Big{[}
	[c,K_\epsilon G_\epsilon^{n-1}] B
	\sigma_{\rm L}^{\rm a}\frac{1}{1-K_\epsilon}\sigma_{\rm R}^{\rm a}
	\partial c c
	\sigma_{\rm L}^{\rm a}\frac{1}{1-K_\epsilon}\sigma_{\rm R}^{\rm a}
	\partial c \frac{1}{-K_\epsilon} G_\epsilon^{-(n-1)}
\Big{]}\notag \\
+
\epsilon &{\rm Tr}\Big{[} 
	c B
	\sigma_{\rm L}^{\rm a}\frac{1}{1-K_\epsilon}\sigma_{\rm R}^{\rm a}
	\partial c \frac{1}{-K_\epsilon} G_\epsilon^{-(n-1)}[c,K_\epsilon G_\epsilon^{n-1}]
	\sigma_{\rm L}^{\rm a}\frac{1}{1-K_\epsilon}\sigma_{\rm R}^{\rm a}
	\partial c 
\Big{]}\notag\\
-
\epsilon &{\rm Tr}\Big{[} 
	c B
	\sigma_{\rm L}^{\rm a}\frac{1}{1-K_\epsilon}\sigma_{\rm R}^{\rm a}
	\partial c c
	\sigma_{\rm L}^{\rm a}\frac{1}{1-K_\epsilon}\sigma_{\rm R}^{\rm a}
	\partial c 
 \Big{]}.
\label{dsdsdsds}
\end{align}
Since $B$ commutes or anti-commutes with BCCOs, $f(K)$ and $[c, f(K)]$,  
the form of the first term in \eqref{dsdsdsds} is something like ${\rm Tr}[B \varphi]$,
where $\varphi$ is some string field commuting with $B$.
Then we can show that this term vanishes as follows: 
\begin{align}
{\rm Tr}[B \varphi]
=
{\rm Tr}[BcB \varphi]
=
{\rm Tr}[B^2c \varphi]
=
0,
\label{B0}
\end{align} 
where in the first equation we use $B = BcB$ and in the next we use $[B, \varphi]=0$ and also the cyclicity of $\rm Tr$.
In the remaining terms of \eqref{dsdsdsds}, 
the number of BCCOs can be reduced by using $[\sigma_{\rm L,R}^{\rm a}, {\rm ghosts}]=0$, 
the cyclicity of $\rm Tr$, and $ \sigma_{\rm R}^{\rm a} \sigma_{\rm L}^{\rm a} = 1$. 
Then, the number of the BCCOs on the RHS of \eqref{eoms} becomes two.
The contribution of the two BCCOs always reduces to the following CFT correlator in the matter sector:
\begin{equation}
\langle \sigma_{\rm L}^{\rm a}(0)\sigma_{\rm R}^{\rm a}(s_1)\rangle_{{\rm C}_{s_1+s_2}}^{\rm ma}
=
g_{\rm a}.
\end{equation}
Here $\sigma_{\rm R,L}^{\rm a}(s)$ express the BCCOs in CFT corresponding to the string fields $\sigma_{\rm R,L}^{\rm a}$, 
and $\langle \cdot \rangle_{{\rm C}_L}$ is the correlator of CFT 
on a semi-infinite cylinder of circumference $L$ in the sliver frame. 
Since the correlator is independent of the positions of BCCOs, we have 
\begin{equation}
\langle \partial \sigma_{\rm L}^{\rm a}(0)\sigma_{\rm R}^{\rm a}(s_1)\rangle_{{\rm C}_{s_1+s_2}}^{\rm ma}
=
0.
\label{<dss>}
\end{equation}
Note that a commutator among $\sigma_{\rm L}^{\rm a}$ and the function of $K$ 
gives the derivative $\partial \sigma_{\rm L}^{\rm a} = [K, \sigma_{\rm L}^{\rm a}]$, 
and hence in {\rm Tr} with a couple of $\sigma^{\rm a}_{\rm L}$ and $\sigma^{\rm a}_{\rm R}$, it can be set to $0$ because of \eqref{<dss>}.
Then we can move the positions of $\sigma_{\rm L}^{\rm a}$ to the immediate left of $\sigma_{\rm R}^{\rm a}$.
Then we can extract the factor $g_{\rm a}=\sigma_{\rm L}^{\rm a}\sigma_{\rm R}^{\rm a}$ on the RHS 
with replacing $\sigma^{\rm a}_{\rm L, R}$ by $1$:
\begin{align}
& {\rm EOMS}\big{(}[\Psi_{{}^{\rm EM}{}^{+n}}^{\rm a}]_\epsilon\big{)}
-{\rm EOMS}\big{(}[\Psi_n]_\epsilon\big{)} \notag \\
& \qquad =  
g_{\rm a}\bigg{[}
	{\rm Tr}\Big{[} 
		( [\Psi_n]_\epsilon + [\Phi_n^{\rm a}]_\epsilon )  
		\epsilon {\partial \over \partial B} ( [\Psi_n]_\epsilon +  [\Phi_{n}^{\rm a}]_\epsilon) 
	\Big{]}
	-
	{\rm Tr} \Big{[} 
		[\Psi_n]_\epsilon \epsilon {\partial \over \partial B} [\Psi_n]_\epsilon 
	\Big{]}
\bigg{]}\Bigg{|}_{\sigma_{\rm L,R}^{\rm a}=1}\notag\\
& \qquad =
g_{\rm a}
\big{[}
	{\rm EOMS}\big{(}[\Psi_{n+1}]_\epsilon\big{)}-{\rm EOMS}\big{(}[\Psi_{n}]_\epsilon\big{)}
\big{]},
\label{eoms2}
\end{align}
where in the last expression
we have used the relation 
$
[ \Psi_n + \Phi_n^{\rm a} ]|_{\sigma_{\rm L,R}^{\rm a}=1}=
\Psi_{{}^{\rm EM}{}^{+n}}^{\rm a}|_{\sigma_{\rm L,R}^{\rm a}=1}
=
\Psi_{n+1}$.
Finlay, the EOM in the strong sense for $\Psi_{{}^{\rm EM}{}^{+n}}^{\rm a}$ is 
\begin{equation}
\lim_{\epsilon \to 0} {\rm EOMS}
\big(
	[\Psi_{{}^{\rm EM}{}^{+n}}^{\rm a}]_\epsilon
\big)
=
\lim_{\epsilon \to 0} 
\bigg[
	(1 - g_{\rm a} ) 
	{\rm EOMS}
	\big(
		[\Psi_n]_\epsilon
	\big)
	+
	g_{\rm a}
	{\rm EOMS}\big( [\Psi_{n+1}]_\epsilon \big)
\bigg].
\label{formula}
\end{equation}
Using the result of \cite{Hata:2012cy}\footnote{
See also the paper \cite{Murata:2011ex}.
}:
\begin{equation}
\lim_{\epsilon \to 0} {\rm EOMS}\big( [\Psi_n]_\epsilon \big) 
= 
- {n(n-1) \over \pi} {\rm Im}[{}_1 F_1(2-n,2,2 \pi i)],
\label{1f1}
\end{equation}
we find that $\Psi_{{}^{\rm EM}{}^{+n}}^{\rm a} $ 
with $n=1$ satisfies EOM in the strong sense,
while for $n>1$ it does not for general $g_{\rm a}$.%
\footnote{In the paper \cite{Hata:2012cy}, the generic result \eqref{1f1} is derived by using the ``$s$-$z$ trick''.
In the appendix \ref{cal}, we show a calculation in the specific case $ \lim_{\epsilon \to 0}{\rm EOMS}
\big{(}
	[\Psi_{{}^{\rm EM}{}^{+2}}^{\rm a}]_\epsilon
\big{)}$ without using it.
}
We also find that there is a special value of $g_{\rm a}$ for each $n$ 
for which $ \Psi_{{}^{\rm EM}{}^{+n}}^{\rm a} $ satisfies the EOM in the strong sense.
Possible interpretations of these solutions are left for the future work. 

\subsection{Energy of the Solution}
Next we check the energy of the solution $\Psi_{{}^{\rm EM}{}^{+1}}^{\rm a}$.
It can be easily evaluated as 
\begin{align}
\lim_{\epsilon\to 0}
E\big{(}[\Psi_{{}^{\rm EM}{}^{+1}}^{\rm a}]_\epsilon \big{)}
&=
-\lim_{\epsilon\to 0}\frac{1}{6}{\rm Tr}
\Big{[}
	{[\Psi_{{}^{\rm EM}{}^{+1}}^{\rm a}]_{\epsilon}}^{3}
\Big{]}\notag\\
& =
-\lim_{\epsilon\to 0}\frac{1}{6}{\rm Tr}
\Big{[}
	\big{[} U \Sigma_{\rm L}^{\rm a}
		(- \Psi_0) 
		\Sigma_{\rm R}^{\rm a} U^{-1} 
	\big{]}_\epsilon{}^3 
\Big{]} \notag \\
&=
\frac{1}{6}{\rm Tr}
\Big{[}
	\Sigma_{\rm L}^{\rm a} \Psi_0 {}^3 \Sigma_{\rm R}^{\rm a}
\Big{]}\notag \\
&=
\frac{1}{6} g_{\rm a} {\rm Tr}
\Big{[}
	\Sigma_{\rm L}^{\rm a} \Psi_0 {}^3 \Sigma_{\rm R}^{\rm a}
\Big{]}\Big|_{\sigma_{\rm L,R}^{\rm a} =1 } \notag \\
&=
\frac{1}{2\pi^{2}}g_{\rm a} \notag \\
& = 
E(\Psi_0) +{g_{\rm a} \over 2 \pi^2 } +{1 \over 2 \pi^2}
.
\label{energy_em+1}
\end{align} 
In the first line we used the EOM in the strong sense,  
and in the next line, the second expression for $\Psi_{{}^{\rm EM}{}^{+1}}^{\rm a}$ in the equation \eqref{a+1} is used.
Then the gauge parameters $U$ and $U^{-1}$ cancel by using the cyclicity of ${\rm Tr}$.
We further use the relation $\Sigma_{\rm R}^{\rm a} \Sigma_{\rm L}^{\rm a} = 1$
to obtain the third line.
Note that we do not need the regularization parameter $\epsilon$ to define $\Sigma_{\rm L}^{\rm a} \Psi_0{}^3 \Sigma_{\rm R}^{\rm a}$ 
since the singularity occurs only through $U^{-1}$. 
Since the third expression includes two BCCOs,
we can replace both BCCOs with identity string fields $1$
and multiply the factor $g_{\rm a}$, as explained after \eqref{<dss>}.

Recalling that the energy of $\Psi_{{}^{\rm EM}}^{\rm a}$ is $ E(\Psi_0) + g_{\rm a} / 2\pi^2 $,
we realize that the energy of the solution is increased by one unit of the energy of the D25-brane, $1/2 \pi^2$, 
through the gauge transformation $U^{-1}$.
So the solution $\Psi_{{}^{\rm EM}{}^{+1}}^{\rm a}$
may be interpreted as the multiple-brane solution which includes 
the D-brane described by the EM solution 
$\Psi_{{}^{\rm EM}}^{\rm a}$ plus the D25-brane.

\subsection{Extension to $\Psi_{{}^{\rm EM}{}^{+n}}^{\rm a+b}$}
Next we consider the singular gauge transformation for the EM solution which is described by
BCFT$_{\rm a}$ and BCFT$_{\rm b}$:
\begin{equation}
\Psi_{{}^{\rm EM}{}^{+n}}^{\rm a+b} 
=
U^{n}(Q+\Psi_{{}^{\rm EM}}^{\rm a+b})U^{-n} 
=
\Psi_n+\Phi_{n}^{\rm a}+\Phi_{n}^{\rm b}. 
\end{equation}
Recall that the EM solution $\Psi_{{}^{\rm EM}}^{\rm a+b}$ is defined by \eqref{Psi-EM-a+b}
and we assume the orthogonality of BCCOs \eqref{sigmaab}. 

The EOM in the strong sense for $\Psi_{{}^{\rm EM}{}^{+n}}^{\rm a+b}$ is as follows:
\begin{align}
& {\rm EOMS}\big{(}[\Psi_{{}^{\rm EM}{}^{+n}}^{\rm a+b}]_\epsilon\big{)} \notag \\
& \qquad =  
{\rm Tr}\Big{[} 
	(
		 [\Psi_n]_\epsilon 
		+
		 [\Phi_n^{\rm a}]_\epsilon 
		+
		 [\Phi_n^{\rm b}]_\epsilon
	)  
	\epsilon {\partial \over \partial B} 
	(
		 [\Psi_n]_\epsilon 
		+
		 [\Phi_{n}^{\rm a}]_\epsilon 
		+ 
		[\Phi_n^{\rm b}]_\epsilon
	)
 \Big{]}.
\label{eomab}
\end{align}
As for the terms with four BCCOs, 
there are cross terms among $\Phi_n^{\rm a}$ and $\Phi_n^{\rm b}$
in addition to the terms like \eqref{EOMiSSa+n}.
The explicit form of these terms can be obtained from \eqref{dsdsdsds} simply by replacing 
second $\sigma_{\rm L,R}^{\rm a}$ in each term of \eqref{dsdsdsds} with $\sigma_{\rm L,R}^{\rm b}$:
\begin{align}
&\hspace{-1.0cm} {\rm Tr}\Big{[}
	[\Phi_n^{\rm a}]_\epsilon 
	\big{(}
		\epsilon\frac{\partial}{\partial B} [\Phi_n^{\rm b}]_\epsilon
	\big{)}
\Big{]}\notag\\
=
& -\epsilon {\rm Tr}
\Big{[}
 	[c,K_\epsilon G_\epsilon^{n-1}] B
	\sigma_{\rm L}^{\rm a}\frac{1}{1-K_\epsilon}\sigma_{\rm R}^{\rm a}
	\partial c \frac{1}{-K_\epsilon} G_\epsilon^{-(n-1)} \notag\\
&\qquad\qquad
	\times [c,K_\epsilon G_\epsilon^{n-1}]
	\sigma_{\rm L}^{\rm b}\frac{1}{1-K_\epsilon}\sigma_{\rm R}^{\rm b}
	\partial c \frac{1}{-K_\epsilon} G_\epsilon^{-(n-1)}
 \Big{]}\notag\\
& +
\epsilon {\rm Tr}
\Big{[}
	[c,K_\epsilon G_\epsilon^{n-1}] B
	\sigma_{\rm L}^{\rm a}\frac{1}{1-K_\epsilon}
	(
		\sigma_{\rm R}^{\rm a}\sigma_{\rm L}^{\rm b}
	)
	\partial c c
	\frac{1}{1-K_\epsilon}\sigma_{\rm R}^{\rm b}
	\partial c \frac{1}{-K_\epsilon} G_\epsilon^{-(n-1)}
 \Big{]}\notag \\
& +\epsilon {\rm Tr}\Big{[}
	c B
	(\sigma_{\rm R}^{\rm b} \sigma_{\rm L}^{\rm a})
	\frac{1}{1-K_\epsilon}\sigma_{\rm R}^{\rm a}
	\partial c \frac{1}{-K_\epsilon} G_\epsilon^{-(n-1)}
 	[c,K_\epsilon G_\epsilon^{n-1}]
	\sigma_{\rm L}^{\rm b}\frac{1}{1-K_\epsilon}
	\partial c 
 \Big{]}\notag\\
& -
\epsilon {\rm Tr}
\Big{[} 
	c B
	(\sigma_{\rm R}^{\rm b}\sigma_{\rm L}^{\rm a})
	\frac{1}{1-K_\epsilon}(\sigma_{\rm R}^{\rm a} \sigma_{\rm L}^{\rm b})
	\partial c c
	\frac{1}{1-K_\epsilon}
	\partial c 
 \Big{]}.
\end{align}
Here, in the third and the fourth terms we used the cyclicity of $\rm Tr$ to shift the positions of BCCOs.
Then we use the equation \eqref{B0} in the first term and the orthogonality of BCCOs in the remaining terms
to obtain 
\begin{align}
{\rm Tr}\Big{[}
	[\Phi_n^{\rm a}]_\epsilon
	\big{(}
		\epsilon\frac{\partial}{\partial B} [\Phi_n^{\rm b}]_\epsilon
	\big{)}
\Big{]}
=
0.
\end{align}
The same result can be also derived by using the orthogonality $\Sigma_{\rm L,R}^{\rm a} \Sigma_{\rm R,L}^{\rm b} =0$ 
($a \neq b$) for 
\begin{equation}
{\rm Tr}
\Big{[}
	[\Phi_n^{\rm a}]_\epsilon
	\big{(}
		\epsilon\frac{\partial}{\partial B} [\Phi_n^{\rm b}]_\epsilon
	\big{)}
\Big]
=
{\rm Tr}
\Big[
	[ U^n \Sigma_{\rm L}^{\rm a} \Psi_0 \Sigma_{\rm R}^{\rm a} U^{-n} ]_\epsilon
	\big(
		\epsilon {\partial \over \partial B}
		[ U^n \Sigma_{\rm L}^{\rm b} \Psi_0 \Sigma_{\rm R}^{\rm b} U^{-n} ]_\epsilon
	\big)
\Big].
\end{equation}
By recombining as $\Psi_{{}^{\rm EM}{}^{+n}}^{\rm a,b} = \Psi_n + \Phi^{\rm a,b}_n$ we have
\begin{align}
& \lim_{\epsilon \to 0} 
{\rm EOMS}
\big{(}
	[\Psi_{{}^{\rm EM}{}^{+n}}^{\rm a+b}]_\epsilon
\big{)} \notag \\
& \qquad =
\lim_{\epsilon \to 0} 
\bigg[
	{\rm EOMS}\big{(} [\Psi_{{}^{\rm EM}{}^{+n}}^{\rm a}]_\epsilon \big{)}
	+
	{\rm EOMS}\big{(} [\Psi_{{}^{\rm EM}{}^{+n}}^{\rm b}]_\epsilon \big{)}
	-
	{\rm EOMS}\big{(} [\Psi_{n}]_\epsilon \big{)}
\bigg] 
\notag \\
& \qquad =
\lim_{\epsilon \to 0} 
\bigg[
	\big(1 - ( g_{\rm a} + g_{\rm b}) \big)
	{\rm EOMS}
	\big(
		[\Psi_n]_\epsilon
	\big)
	+
	( g_{\rm a} + g_{\rm b} ) 
	{\rm EOMS}\big( [\Psi_{n+1}]_\epsilon \big)
\bigg].
\end{align}
We find again that, for generic $g_{\rm a,b}$, 
only $\Psi_{{}^{\rm EM}{}^{+1}}^{\rm a+b}$ satisfies 
the EOM in the strong sense.
The energy of this solution is
\begin{align}
\lim_{\epsilon \to 0} E
\big{(}
	[\Psi_{{}^{\rm EM}{}^{+1}}^{\rm a+b}]_\epsilon
\big{)}
=
-\frac{1}{6}\lim_{\epsilon \to 0}{\rm Tr}
\Big{[}
	{[\Phi_1^{\rm a}+\Phi_1^{\rm b}]_\epsilon}^3
\Big{]}.
\end{align}
Here the cross terms vanish as
\begin{align}
\Phi^{\rm a}_n\Phi^{\rm b}_n
&=
U^{n}\Sigma_{\rm L}^{\rm a}(-\Psi_0)\Sigma_{\rm R}^{\rm a}U^{-n}
U^{n}\Sigma_{\rm L}^{\rm b}(-\Psi_0)\Sigma_{\rm R}^{\rm b}U^{-n}\notag\\
&=
U^{n}\Sigma_{\rm L}^{\rm a}\Psi_0
(
	\Sigma_{\rm R}^{\rm a} \Sigma_{\rm L}^{\rm b}
)
\Psi_0\Sigma_{\rm R}^{\rm b}U^{-n}\notag \\
&=
0.
\end{align}
Therefore, the energy is given by the summation of the energy of the two solutions:
\begin{align}
\lim_{\epsilon \to 0} E
\big{(}
	[\Psi_{{}^{\rm EM}{}^{+1}}^{\rm a+b}]_\epsilon
\big{)}
&=
\lim_{\epsilon \to 0} E
\big{(}
	[\Psi_{{}^{\rm EM}{}^{+1}}^{\rm a}]_\epsilon
\big{)}
+
\lim_{\epsilon \to 0} E
\big{(}
[\Psi_{{}^{\rm EM}{}^{+1}}^{\rm b}]_\epsilon
\big{)}\notag\\
&=
\frac{1}{2\pi^2}(g_{\rm a}+g_{\rm b}) \notag \\
& =
E(\Psi_0) + \frac{g_{\rm a}}{2\pi^2} + { g_{\rm b} \over 2 \pi^2 } + {1 \over 2 \pi^2 }
.
\end{align}

Further extension (${\rm a}+{\rm b}+ {\rm c}+\ldots$) is straightforward.

\section{An Example: D24+D25-brane}
Finally, as a further support for our observation 
that the singular gauge transformation acting on 
the EM solution produces an additional D25-brane, 
we consider a concrete solution and study 
the profile of the tachyon field \cite{Erler:2014eqa}. 
After using the $K_{\epsilon}$ regularization, the solution we consider is given by
\begin{equation}
[\Psi_{{}^{\rm EM}{}^{+1}}^{{}_{\rm ND}}]_{\epsilon}
=
cBK_{\epsilon}
\sigma_{\rm L}^{\rm ND}\frac{1}{1-K_{\epsilon}}\sigma_{\rm R}^{\rm ND}
K_{\epsilon}c\frac{1}{-K_{\epsilon}},
\label{ND+1}
\end{equation}
where $\sigma_{\rm L,R}^{\rm ND}$ are the BCCOs which change the Neumann boundary condition of $X^1$
to the Dirichlet boundary condition and vice versa. They satisfy the following algebra \cite{Erler:2014eqa}:
\begin{equation}
\sigma^{\rm ND}_{\rm R} \sigma^{\rm ND}_{\rm L} = 1, \quad 
\sigma^{\rm ND}_{\rm L} \sigma^{\rm ND}_{\rm R} = g_{\rm ND} = {1 \over R}.
\end{equation}
Let us assume that $X^{1}$ is compactified as $X^{1} \simeq X^{1}+2\pi R$, and that the end points of the string
is at $X^1 = 0$.

We calculate the tachyon profile as in \cite{Erler:2014eqa}.
The tachyon field $T(X^1)$ is expanded as
\begin{equation}
T(X^1)
=
\sum_{n \in \mathbb{Z}} t_n e^{i\frac{n}{R}X^1}.
\end{equation}
The coefficient $t_n$ can be computed by using the state $|\tilde{T}_n\rangle$ which is dual to 
 the tachyon state $|T_n\rangle$ satisfying ${\rm Tr}[\tilde{T}_nT_m]=\delta_{n,m}$:
\begin{align}
|T_n\rangle&
=
ce^{i\frac{n}{R}X^1}(0)|0\rangle, \\
|\tilde{T}_n\rangle 
&=
-\frac{1}{2\pi R}c \partial c e^{-i\frac{n}{R}X^1}(0)|0\rangle.
\end{align}
Here, the state $|0\rangle$ is the $SL(2,\mathbb{R})$-invariant vacuum
which is defined on the unit semi-circle of the upper half-plane (UHP),
while the vertex operators are inserted at the origin.
Then the coefficient $t_n$ for \eqref{ND+1} is given by
\begin{align}
t_n
&=\lim_{\epsilon \to 0}{\rm Tr}
\Big{[}
	\tilde{T}_n
	[\Psi_{{}^{\rm EM}{}^{+1}}^{{}_{\rm ND}}]_{\epsilon}
\Big{]}\notag\\
&=
-
\lim_{\epsilon \to 0}\frac{\pi}{2}
\int_{0}^{\infty}dx_{1}
\int_{0}^{\infty}dz_{1}
\lim_{y_{1},y_{2}\to 0}
\partial_{y_{1}}\partial_{y_{2}}\,
e^{-\epsilon(z_{1}+y_{1}+y_{2})}
e^{-(1+\epsilon)x_1}\notag\\
&~~~
\times\Big{\langle} 
\int^{-i \infty}_{i \infty}
\frac{dw}{2\pi i}
b(w) c(0) c\partial c(z_1+\frac{1}{2})c(z_1+1)
\Big{\rangle}_{{\rm C}_{1+x_1+y_1+y_2+z_1}}^{\rm gh}
\notag\\
&~~~\times
\Big{\langle}
	\frac{1}{2\pi R} f_{\rm s}\circ e^{-i\frac{n}{R} X^1}(0)
	\sigma_{\rm L}^{\rm ND}(\frac{1}{2}+y_1)
	\sigma_{\rm R}^{\rm ND}(\frac{1}{2}+y_1+x_1)
\Big{\rangle}_{{\rm C}_{1+x_1+y_1+y_2+z_1}}^{\rm ma},
\end{align}
where $f_{\rm s}$ is the conformal transformation which maps UHP to the sliver frame: $f_{\rm s}(\xi)=(2/\pi)\arctan \xi.$
Here we used the equations \eqref{x}-\eqref{z}.
Using the following correlators \cite{Schnabl:2005gv,Okawa:2006vm}\cite{Erler:2014eqa, Kiermaier:2010cf,Mukhopadhyay:2001ey}, 
we can compute $t_n$:
\begin{figure}[t]
\begin{center}
\includegraphics[width=13cm]{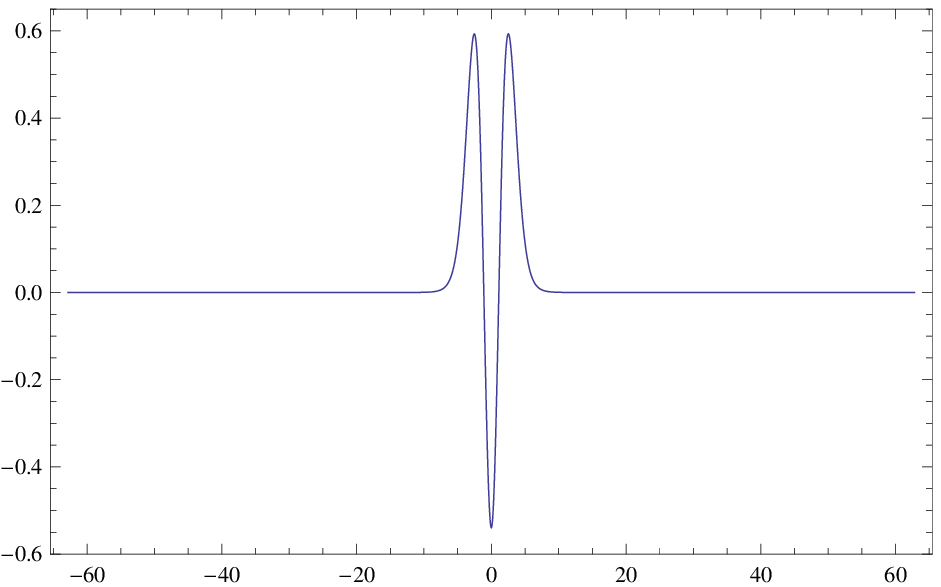}
\put(-185,-12){ $X^1$}
\put(-410,113){ $T(X^1)$}
\caption{The profile of the tachyon field is shown. 
The compactification radius is taken to be $R$=20.
We numerically computed by setting 
$|n| \leq 100$ and $\epsilon=0.001$.}
\label{fig:1}
\end{center}
\end{figure}
\begin{align}
&
\Big{\langle}
	\int^{-i \infty}_{i \infty} \frac{dw}{2\pi i}
	b(w)c(0)c(s_1)c(s_1+s_2)c(s_1+s_2+s_3)
\Big{\rangle}_{{\rm C}_{s_1+s_2+s_3+s_4}}^{\rm gh} \notag \\
&\qquad \qquad =
-\frac{L^2}{4\pi^3}
\big{(}
	s_3\sin 2\theta_{s_1}
	-
	(s_2+s_3)\sin2\theta_{s_1+s_2}
	+
	s_2\sin2\theta_{s_1+s_2+s_3}\notag\\
& \qquad \qquad \qquad 
	+
	s_1\sin2\theta_{s_3}
	-
	(s_1+s_2)\sin2\theta_{s_2+s_3}
	+
	(s_1+s_2+s_3)\sin2\theta_{s_2}
\big{)},
\label{gh4}\\[2mm]
&\Big{\langle}
	\frac{1}{2\pi R}(f_{\rm s}\circ  e^{-i\frac{n}{R}X^1})(0)
	\sigma_{\rm L}^{\rm ND}(s_1)
	\sigma_{\rm R}^{\rm ND}(s_1+s_2)
\Big{\rangle}_{{\rm C}_{s_1+s_2+s_3}}^{\rm ma} \notag \\
& \qquad \qquad =
\frac{2^{-2(n/R)^2}}{R}
\Big{(}
	\frac{2\sin\theta_{s_2}}{L\sin\theta_{s_1}\sin\theta_{s_1+s_2}}
\Big{)}^{(n/R)^2}.
\label{ma3}
\end{align}
Here, $\theta_{s_i}\equiv \pi {s_i}/L$ and $L$ is the circumference of the cylinder, i.e. $L=\sum_{i=1}^4 s_i$ in \eqref{gh4} and $L=\sum_{i=1}^3 s_i$ in \eqref{ma3}.

The figure 1 shows the numerical result for the profile of the tachyon field.
Recall that in the case of the EM solution $\Psi_{{}^{\rm EM}}^{{}_{\rm ND}}$,
far away from $X^1=0$, the value of the tachyon field 
asymptotically approaches to that for the tachyon vacuum solution $\Psi_0$ \cite{Erler:2014eqa}.
In our case of $\Psi_{{}^{\rm EM}{}^{+1}}^{{}_{\rm ND}}$, the tachyon field  asymptotically approaches to zero,
i.e. the value of the perturbative vacuum $\Psi_{1}=0$ representing the D25-brane.
Therefore we interpret the solution $\Psi_{{}^{\rm EM}{}^{+1}}^{{}_{\rm ND}}$
to describe a multiple-brane solution in which the D24-brane is placed around $X^1=0$ on the D25-brane.
Indeed, from \eqref{energy_em+1} with $g_{\rm ND}=1/R$ \cite{Erler:2014eqa},
the energy of the solution \eqref{ND+1} is given by
\begin{align}
\lim_{\epsilon \to 0} E
\big{(}
	[\Psi_{{}^{\rm EM}{}^{+1}}^{{}_{\rm ND}}]_\epsilon
\big{)}
=
\frac{1}{2\pi^2}\frac{1}{R}
=
E(\Psi_0)+T_{25}+
{T_{\rm 24 }V_{24} \over V_{25}} .
\end{align}
Here, $T_{\rm 25} = 1 / 2 \pi^2$ and $T_{\rm 24} = T_{\rm 25} \times 2 \pi$ 
are the tensions of a D25-brane and a D24-brane, respectively. 
The volumes  $V_{24}$ and $V_{25}$ are the spacial volumes of indicated dimensions
and they are related by $V_{25} = V_{24} \times 2 \pi R$.
Then the result coincides with the summation of the energy of a D25-brane and a D24-brane 
multiplied by the present normalization factor $1/V_{25}$.
This result provides a support for the above observation of multiple-brane solution.

\section{Conclusion}
In this paper, we have studied candidates for 
multiple-brane solutions of bosonic open string field theory.
To construct them we performed the singular gauge transformations for the solution by Erler and Maccaferri 
which includes the boundary condition changing operators.
Since our candidates include the singular string field $1/K$, 
we adopted the $K_\epsilon$ regularization.
Then we checked the EOM in the strong sense and found that 
non-vanishing terms are expressed by the non-vanishing terms
in the case without BCCOs
\cite{Murata:2011ex,Hata:2011ke, Hata:2012cy}.
Only the unique candidate 
$\Psi_{{}^{\rm EM}{}^{+1}}^{\rm a}$ satisfies the EOM in the strong sense for a generic value of 
the disk partition function $g_{\rm a}$ in BCFT$_{\rm a}$.
We have also noticed that for each candidate $\Psi_{{}^{\rm EM}{}^{+n}}^{\rm a}$ with $n > 1$,
there exists a special value of $g_{\rm a}$ for which the EOM in the strong sense seems to be accidentally satisfied.
We leave further studies of this phenomena for the future work.
Since there are two BCCOs in our candidates, 
correlators including four BCCOs could have appeared 
in our calculations of the EOM in the strong sense,
and such correlators could have made the computations difficult.
However, in the actual computations they just vanish and we need only correlators with two BCCOs.
Next we have studied the energy of our solution $\Psi_{{}^{\rm EM}{}^{+1}}^{\rm a}$ and found that 
the gauge transformation changes the energy by the tension of a D25-brane.
This result can be regarded as a support for the expectation that 
our solution describes the background which includes two D-branes, i.e. 
the D-brane of the original EM solution and also the D25-brane coming from the singular gauge transformation.
As a further check of the existence of the additional D25-brane,
we chose the EM solution representing a D24-brane 
and have calculated the profile of the tachyon field.
The result agrees with the above expectation.

\section*{Acknowledgements}
The authors would like to thank T.~Kojita and T.~Masuda for useful discussions and comments. 
They are also grateful to members in the particle theory group in CST, Nihon University for discussions and encouragements. 
K.~S. would like to thank the Yukawa Institute for Theoretical Physics at Kyoto University. 
Discussions during the YITP workshop on ``Developments in String Theory and Quantum Field Theory'' (YITP-W-15-12) were useful to complete this work.

\appendix
\section
{A check of the Equation of Motion in the Strong Sense} 
%without Using the $\bm s$-$\bm z$ Trick}
%Detailed calculations of checking the equation of motion 
%in the strong sense 
%${\rm EOMS}
%\big{(}[\Psi_{{}^{\rm EM}{}^{+2}}^{\rm a}{]}_\epsilon \big{)}$
%}
\label{cal}

We give calculation of the EOM in the strong sense for
$
\Psi_{{}^{\rm EM}{}^{+n}}^{\rm a}
$ \eqref{a+n} in the case of $n=2$
without using the $s$-$z$ trick 
which is used in the papers \cite{Murata:2011ex,Hata:2012cy} to derive the generic formula \eqref{1f1}.
%The derivation of \eqref{1f1} .
The EOM in the strong sense \eqref{eomsn} with $n=2$ is expressed as follows
\begin{align}
{\rm EOMS}
\big{(}
	[\Psi_{{}^{\rm EM}{}^{+2}}^{\rm a}]_\epsilon
\big{)}
&=
{\rm EOMS}\big{(}[\Psi_2]_\epsilon\big{)}
+
{\rm Tr}
\Big{[}
	[\Psi_2]_\epsilon(\epsilon {\partial \over \partial B}
	[\Phi_{2}^{\rm a}]_\epsilon)
\Big{]}\notag \\
&~~~~~+{\rm Tr}
\Big{[}
	[\Phi^{\rm a}_2]_\epsilon
	(
		\epsilon {\partial \over \partial B}
		[\Psi_{2}]_\epsilon
	)
\Big{]}
+
{\rm Tr}
\Big{[}
	[\Phi^{\rm a}_2]_\epsilon
	(
		\epsilon {\partial \over \partial B}
		[\Phi_{2}^{\rm a}]_\epsilon
	)
\Big{]}.
\label{eomseoms}
\end{align}
The explicit form of each term of \eqref{eomseoms} is
\begin{align}
&{\rm Tr}
\Big{[}
	[\Psi_2]_\epsilon
		(\epsilon {\partial \over \partial B} 
		[\Phi_2^{\rm a}]_\epsilon
	)
\Big{]}
=
\epsilon {\rm Tr}
\Big{[}
	\sigma_{\rm L}^{\rm a}\frac{1}{1-K_\epsilon}\sigma_{\rm R}^{\rm a} 
	\partial c \frac{1}{-K_\epsilon}G_\epsilon^{-1}
	cK_\epsilon G_\epsilon BcG_\epsilon^{-1}
	cK_\epsilon G_\epsilon
\Big{]},\label{psbph}  \\ 
&{\rm Tr}
\Big{[}
	[\Phi^{\rm a}_2]_\epsilon
	(
		\epsilon {\partial \over \partial B}
		[\Psi_{2}]_\epsilon
	)
\Big{]}
=
{\rm Tr}
\Big{[}
	[\Psi_2]_\epsilon
	(
		\epsilon {\partial \over \partial B}
		[\Phi_2^{\rm a}]_\epsilon
	)
\Big{]}
-
\epsilon{\rm Tr}
\Big{[}
	\sigma_{\rm L}^{\rm a}\frac{1}{1-K_\epsilon}\sigma_{\rm R}^{\rm a}
	c \partial c G_\epsilon^{-1}
	cK_\epsilon G_\epsilon
\Big{]},
\end{align}
and
\begin{align}
{\rm Tr}
\Big{[}
	[\Phi^{\rm a}_2]_\epsilon
	(
		\epsilon {\partial \over \partial B}
		[\Phi_{2}^{\rm a}]_\epsilon
	)
\Big{]} 
=
&
-2\epsilon {\rm Tr}
\Big{[}
	Bc \partial c \frac{1}{1-K_\epsilon}
	\sigma_{\rm R}^{\rm a}
	\partial c \frac{1}{-K_\epsilon}G_\epsilon^{-1}
	[c,K_\epsilon G_\epsilon]
	\sigma_{\rm L}^{\rm a}
	\frac{1}{1-K_\epsilon}
\Big{]}\notag\\
&
-\epsilon{\rm Tr}
\Big[
	\sigma_{\rm L}^{\rm a}
	Bc\partial c \frac{1}{1-K_\epsilon}
	c \partial c \frac{1}{1-K_\epsilon}
	\sigma_{\rm R}^{\rm a}
\Big]
.
\label{EOMSa+n}
\end{align}
Here, \eqref{dsdsdsds} is used in \eqref{EOMSa+n}.
By gathering \eqref{psbph}-\eqref{EOMSa+n}, we have
\begin{align}
%&\hspace{-1cm}
{\rm EOMS}
\big{(}
	[\Psi_{{}^{\rm EM}{}^{+2}}^{\rm a}]_\epsilon
\big{)}%\notag\\
=
&{\rm EOMS}\big{(}[\Psi_{2}]_\epsilon\big{)}\\
&+
2\epsilon {\rm Tr}
\Big{[}
	BcG_\epsilon^{-1}cK_\epsilon G_\epsilon
	\sigma_{\rm L}^{\rm a}\frac{1}{1-K_\epsilon}\sigma_{\rm R}^{\rm a} 
	\partial c \frac{1}{-K_\epsilon}G_\epsilon^{-1}
	cK_\epsilon G_\epsilon
\Big{]}\label{eomsa}\\
&-
\epsilon{\rm Tr}
\Big{[}
	Bc \partial c G_\epsilon^{-1}cK_\epsilon G_\epsilon
	\sigma_{\rm L}^{\rm a}\frac{1}{1-K_\epsilon}\sigma_{\rm R}^{\rm a}
	c
\Big{]}
\label{eomsb}\\
&-
2\epsilon {\rm Tr}
\Big{[}
	Bc \partial c \frac{1}{1-K_\epsilon}\sigma_{\rm R}^{\rm a}
	\partial c \frac{1}{-K_\epsilon}G_\epsilon^{-1}
	[c,K_\epsilon G_\epsilon]
	\sigma_{\rm L}^{\rm a}\frac{1}{1-K_\epsilon}
\Big{]}
\label{eomsc}\\
&-
\epsilon{\rm Tr}
\Big{[}
	\sigma_{\rm L}^{\rm a}
	Bc\partial c \frac{1}{1-K_\epsilon}
	c \partial c \frac{1}{1-K_\epsilon}
	\sigma_{\rm R}^{\rm a}
\Big{]}
\label{eomsd}
.
\end{align}
We arrange the position of each $K_\epsilon$ 
in the numerator so that it always appears 
in the form $\partial c = [ K_\epsilon , c]$:
\begin{align}
\eqref{eomsa}
=
g_{\rm a}\Big{(}
&
2\epsilon {\rm Tr}
\Big{[}
	Bc\partial c  \frac{1}{1-K_\epsilon} \partial c
	\frac{1}{K_\epsilon{^2}}\partial c \frac{1}{1-K_\epsilon}
\Big{]}
\notag \\
-&2\epsilon {\rm Tr}
\Big{[}
	Bc\partial c \frac{1}{1-K_\epsilon} \frac{1}{1-K_\epsilon}
	\partial c \frac{1}{K_\epsilon{}^2}\partial c \frac{1}{1-K_\epsilon}
\Big{]}
\notag\\
+&2\epsilon {\rm Tr}
\Big{[}
	Bc\partial c  \frac{1}{1-K_\epsilon} 
	\partial c \frac{1}{-K_\epsilon}\partial c \frac{1}{-K_\epsilon}
\Big{]}
\notag \\
-&2\epsilon {\rm Tr}
\Big{[}
	Bc\partial c \frac{1}{1-K_\epsilon} \frac{1}{1-K_\epsilon} 
	\partial c \frac{1}{-K_\epsilon}\partial c \frac{1}{-K_\epsilon}
\Big{]}
\notag \\
-&2\epsilon {\rm Tr}
\Big{[}
	Bc\frac{1}{-K_\epsilon}\partial c  \frac{1}{1-K_\epsilon}
	c\partial c 
\Big{]}
\notag \\
+&2\epsilon {\rm Tr}
\Big{[}
	Bc\frac{1}{-K_\epsilon}\partial c \frac{1}{1-K_\epsilon}
	\frac{1}{1-K_\epsilon} c\partial c 
\Big{]}
\Big{)},
\label{eomsa2}
\end{align}
where we extract the factor $g_{\rm a}$ because all terms contain two BCCOs, as explained around \eqref{<dss>}.
We define
\begin{align}
{\rm Bcddd}[s_1,s_2,s_3,s_4]
&\equiv
{\rm Tr}\Big{[}Bc\Omega^{s_1}\partial c \Omega^{s_2} \partial c \Omega^{s_3}\partial c \Omega^{s_4}\Big{]}
,\\
{\rm Bcdcd}[s_1,s_2,s_3,s_4]
&\equiv
{\rm Tr}\Big{[}Bc\Omega^{s_1}\partial c \Omega^{s_2} c \Omega^{s_3}\partial c \Omega^{s_4}\Big{]},\\
{\rm Bcddc}[s_1,s_2,s_3,s_4]
&\equiv
{\rm Tr}\Big{[}Bc\Omega^{s_1}\partial c \Omega^{s_2} \partial c \Omega^{s_3} c \Omega^{s_4}\Big{]}.
\end{align}
For simplicity,
we use the certain letters $x_i$ and $z_i,$ as Schwinger parameters corresponding to the following Laplace transformations:
\begin{align}
\frac{1}{1-K_\epsilon}
&=
\int_0^\infty dx_i\,
e^{-(1+\epsilon)x_i}\Omega^{x_i},\notag\\
\frac{1}{-K_\epsilon}
&=
\int_0^\infty dz_i\,
e^{-\epsilon z_i} \Omega^{z_i}.
\end{align}
In the following we omit $\int_0^\infty dx_i$ and $\int_0^\infty dz_i$ and also the exponential factors, e.g.
we abbreviate the term 
\begin{align}
&{\rm Tr}
\Big{[}
	Bc\partial c \frac{1}{1-K_\epsilon} \partial c 
	\frac{1}{K_\epsilon{}^2} \partial c \frac{1}{1-K_\epsilon}
\Big{]}
\notag\\
&\quad =
\int_0^\infty dx_1\int_0^\infty dz_1\int_0^\infty dz_2\,
e^{-(1+\epsilon)x_1-\epsilon(z_1+z_2)}
{\rm Bcddd}[0,x_1,z_1+z_2,x_2]
\end{align}
as 
\begin{equation}
{\rm Bcddd}[0,x_1,z_1+z_2,x_2].
\end{equation}
By using this notation, \eqref{eomsa2} can be written as 
\begin{align}
\eqref{eomsa2}/{g_{\rm a}}
=&
2\epsilon {\rm Bcddd}[0,x_1,z_1+z_2,x_2]\label{a21}\\
&-2\epsilon {\rm Bcddd}[0,x_1+x_2,z_1+z_2,x_3]\label{a22}\\
&+2\epsilon {\rm Bcddd}[0,x_1,z_1,z_2]\label{a23}\\
&-2\epsilon{\rm Bcddd}[0,x_1+x_2,z_1,z_2]\label{a24}\\
&-2\epsilon {\rm Bcdcd}[z_1,x_1,0,0]\label{a25}\\
&+2\epsilon {\rm Bcdcd}[z_1,x_1+x_2,0,0]\label{a26}.
\end{align}
Using the follwing formulae:
\begin{align}
{\rm Bcddd}[s_1,s_2,s_3,s_4]
&=
-\frac{1}{\pi}
(
	\sin2\theta_{s_2}
	+\sin2\theta_{s_3}
	-\sin2\theta_{s_2+s_3}
),\\
{\rm Bcdcd}[s_1,s_2,s_3,s_4]
&=
\frac{L}{2\pi^2}
(
	-\cos2\theta_{s_1}
	+\cos2\theta_{s_2}
	-\cos2\theta_{s_3}
	+\cos2\theta_{s_1+s_2+s_3}\notag\\
&\qquad\quad
	+2\theta_{s_1+s_2}\sin2\theta_{s_2+s_3}	
),\\
{\rm Bcddc}[s_1,s_2,s_3,s_4]
&=
\frac{-L}{2\pi^2}
(
	-\cos2\theta_{s_1}
	+\cos2\theta_{s_1+s_2}
	-\cos2\theta_{s_3}
	+\cos2\theta_{s_2+s_3}\notag\\
&\qquad\quad
+2\theta_{s_1+s_2+s_3}\sin2\theta_{s_2}
)
,
\end{align}
where $L$ is the circumference of the cylinder 
and
$\theta_{s_i} = \pi s_i/L$,
we have
\begin{align*}
\begin{array}{lll}
\displaystyle \lim_{\epsilon \to 0}\eqref{a21}=0, \quad &
\displaystyle \lim_{\epsilon \to 0}\eqref{a22}=0, \quad & 
\displaystyle \lim_{\epsilon \to 0}\eqref{a23}=-4g_{\rm a},\\[2mm]
\displaystyle \lim_{\epsilon \to 0}\eqref{a24}=8g_{\rm a}, \quad &
\displaystyle \lim_{\epsilon \to 0}\eqref{a25}= -4g_{\rm a}, \quad & 
\displaystyle \lim_{\epsilon \to 0}\eqref{a26}=8g_{\rm a}.
\end{array}
\end{align*}
As an example, we write explicitly the computation of \eqref{a23}:
\begin{align}
& \lim_{\epsilon \to 0}\eqref{a23}/g_{\rm a} \notag \\
&=
-\lim_{\epsilon \to 0}
\frac{2\epsilon}{\pi}\int_0^{\infty} dx_1 \int_0^{\infty} dz_1 \int_0^{\infty} dz_2\,
e^{-(1+\epsilon)x_1-\epsilon(z_1+z_2)}
(
	\sin2\theta_{x_1}
	+
	\sin2\theta_{z_1}
	-
	\sin2\theta_{x_1+z_1}
)
\notag\\
&=
-\lim_{\epsilon \to 0}
\frac{2\epsilon}{\pi}\int_0^{\infty} a^2 da\int_0^{1}dc\int_0^{c}db\,
e^{-a(b+\epsilon)}
(
	\sin2b\pi -\sin2c\pi +\sin2(c-b)\pi
)
\notag\\
&=
2\lim_{\epsilon \to 0}
\epsilon \int_0^\infty da\,
e^{-a\epsilon}
\frac{
	16ae^{-a}\pi^2
	-2a^4
	-4a^2\pi^2
	-8a\pi^2
	}
{
	(a^2+4\pi^2)^2
}
\notag\\
&=
2\lim_{\epsilon \to 0}
\int_0^\infty d\tilde{a}\,
e^{-\tilde{a}}
\frac{
	\epsilon^3 16\tilde{a}e^{-\tilde{a}/\epsilon}\pi^2
	-2\tilde{a}^4
	-4\epsilon^2\tilde{a}^2\pi^2
	-8\epsilon^3\tilde{a}\pi^2
	}
{
	\big(\tilde{a}^2
	+(2\pi\epsilon)^2\big)^2
}\notag\\
&=
-4. \label{example}
\end{align}
Here, we change the variables:
\begin{equation}
x_1=ab, \quad z_1=a(c-b), \quad z_2=a(1-c), \notag
\end{equation}
%(x_1,z_1,z_2)\to(ab,a(c-b),a(1-c)),
\begin{equation}
(a=x_1+z_1+z_2, \quad b=\frac{x_1}{x_1+z_1+z_2}, \quad c=\frac{x_1+z_1}{x_1+z_1+z_2}), \notag
\end{equation}
and $\tilde{a}$ is defined as $\tilde{a}\equiv a\epsilon$.
Computations of other terms can be done similarly.

The remaining terms \eqref{eomsb}-\eqref{eomsd} are rewritten as
%For \eqref{eomsb},
\begin{align}
\eqref{eomsb}/g_{\rm a}=&+\epsilon {\rm Bcddc}[0,z_1,x_1,0]\label{b1}\\
&-\epsilon{\rm Bcddc}[0,z_1,x_1+x_2,0]\label{b2}, \\
\eqref{eomsc}/g_{\rm a}=&-2\epsilon{\rm Bcddd}[0,x_1,z_1+z_2,x_2]\label{c1}\\
&+2\epsilon {\rm Bcddd}[0,x_1,z_1+z_2,x_2+x_3]\label{c2}\\
&-2\epsilon {\rm Bcddd}[0,x_1,z_1,x_2],\label{c3} \\
\eqref{eomsd}/g_{\rm a}=& -\epsilon{\rm Bcdcd}[0,x_1,0,x_2].
\label{d1}
\end{align}
After the similar steps as \eqref{example} we obtain the following results:
\begin{align*}
& \lim_{\epsilon \to 0}\eqref{b1}=2g_{\rm a}, \quad 
\lim_{\epsilon \to 0} \eqref{b2}=-4g_{\rm a}, \\
& \lim_{\epsilon\to 0}\eqref{c1}=0,\quad \lim_{\epsilon\to 0}\eqref{c2}=0,\quad
\lim_{\epsilon\to 0}\eqref{c3}=0, \\
& \lim_{\epsilon \to 0}\eqref{d1}=0.
\end{align*}
Finaly, we have
\begin{equation}
\lim_{\epsilon \to 0}
{\rm EOMS}
\big{(}
	[\Psi_{{}^{\rm EM}{}^{+2}}^{\rm a}]_\epsilon
\big{)}
=
6g_{\rm a},
\end{equation}
and this result agrees with
\eqref{formula} with $n=2$
by using \eqref{1f1}
.
%\newpage

%%%%%%%%%%%%%%%%%%%%%%%%%%%%%%%

\end{document}